\documentclass[a4paper,aps,prb,preprint]{revtex4}
\usepackage{graphicx}
\usepackage{pslatex}
\usepackage{amsmath}
\usepackage{isolatin1}
\begin{document}
\title{Detection of the magneto-structural phase coexistence in MnAs epilayers at a very early stage}
\date{\today}
\author{J. Milano} \email{milano@cab.cnea.gov.ar}
\affiliation {CNEA - Centro At\'omico Bariloche and Instituto Balseiro, UNCu, (R8402AGP) San Carlos de Bariloche, Río Negro, Argentina.}
\author{L. B. Steren}
\affiliation {CNEA - Centro At\'omico Bariloche and Instituto Balseiro, UNCu, (R8402AGP) San Carlos de Bariloche, Río Negro, Argentina.}
\author{A. H. V. Repetto Llamazares}
\affiliation {CNEA - Centro At\'omico Bariloche and Instituto Balseiro, UNCu, (R8402AGP) San Carlos de Bariloche, Río Negro, Argentina.}
\author{V. Garcia}
\affiliation {Institut des NanoSciences de Paris, INSP, Universit\'e
Pierre et Marie Curie-Paris 6, Universit\'e Denis Diderot-Paris 7,
CNRS UMR 7588, Campus Boucicaut, 140 rue de Lourmel, 75015 Paris,
France.}
\author{M. Marangolo}
\affiliation {Institut des NanoSciences de Paris, INSP, Universit\'e
Pierre et Marie Curie-Paris 6, Universit\'e Denis Diderot-Paris 7,
CNRS UMR 7588, Campus Boucicaut, 140 rue de Lourmel, 75015 Paris,
France.}
\author{M. Eddrief}
\affiliation {Institut des NanoSciences de Paris, INSP, Universit\'e
Pierre et Marie Curie-Paris 6, Universit\'e Denis Diderot-Paris 7,
CNRS UMR 7588, Campus Boucicaut, 140 rue de Lourmel, 75015 Paris,
France.}
\author{V. H. Etgens}
\affiliation {Institut des NanoSciences de Paris, INSP, Universit\'e
Pierre et Marie Curie-Paris 6, Universit\'e Denis Diderot-Paris 7,
CNRS UMR 7588, Campus Boucicaut, 140 rue de Lourmel, 75015 Paris,
France.}
\begin{abstract}
We report on the appearance of magnetic stripes in MnAs/GaAs(100) epilayers at temperatures well below the ferromagnetic transition of the system. The study has been performed by ferromagnetic resonance experiments (FMR) on MnAs epilayers grown on (100) and (111) GaAs substrates. The FMR spectra of the MnAs/GaAs(100) samples at 180 K reveal the appearance of zones of different magnetic behavior with respect to the low-temperature homogeneous ferromagnetic phase. The angular and the temperature dependence of the spectra serve us to detect the inter-growth of the non-magnetic phase into the ferromagnetic phase at a very early stage of the process. The experimental data show that the new phase nucleates in a self-arranged array of stripes in MnAs/GaAs(100) thin films while it grows randomly in the same films grown on GaAs(111).
%\pacs{75.70.Ak ; 76.50.+g ; 75.30.Gw}
\end{abstract}
\maketitle

Due to their physical and chemical properties, the MnAs compound is a promising material for spintronic developments \cite{zutic}. Indeed, with room temperature ferromagnetism and extremely low interface reactivity with GaAs, MnAs is a strong candidate for spin injection devices. Since the earlier studies of MnAs epilayers, the coexistence of two magneto-structural phases in a finite temperature
range has been one of the most interesting topic of research.\cite{daweritz} One of these phases is
ferromagnetic ($\alpha$-phase) and the other one is non-magnetic ($\beta$-phase).\cite{wilson} The bulk
compound does not show such phase coexistence; on the contrary, it presents a
magneto-structural first-order transition at 313 K from the low temperature $\alpha$ phase to the $\beta$ one above this temperature \cite{willis,goodenough,blois}.
The origin of the phase coexistence in MnAs thin films has been well addressed and is due to strain
induced by the substrate \cite{adriano,rungger,garciaPRLneutrons}.
%The substrate does not allow
%a free structural contraction from the $\alpha$ to the $\beta$ phase as
%occurs in the bulk compound \cite{kaganerPRL85}.
%The $\alpha$-$\beta$ coexistence arrangement depends on the substrate
%orientation, i.e., over GaAs(100), MnAs thin films show an ordered
%arrangement of stripes-like domains, alternating the $\alpha$ and
%the $\beta$ phases \cite{das}. On the other hand, when the MnAs is
%grown over GaAs(111), the coexistence is more imbricated \cite{takagaki}.
The phase coexistence has been
experimentally observed using several and very different techniques, \cite{garciaPRLneutrons,kaganerPRB66,mattoso,iikawaPRB65,kastner,mohanty,plake,ney,vidal}.
%The phase coexistence has been
%experimentally observed using several and very different techniques, i.e., X-ray and neutron
%diffraction \cite{garciaPRLneutrons,kaganerPRB66,mattoso}, magneto-optical Kerr
%effect \cite{iikawaPRB65}, scanning probe
%microscopies \cite{kastner,mohanty,plake}, ferromagnetic resonance \cite{ney}, and
%optical spectroscopy \cite{vidal}.
All of the experimental works referred previously show that the phase coexistence starts around 273 K. However, in a recent paper \cite{sterenHc}, we observed, above 200 K, an increase of the coercive field with increasing temperature, which is very unusual for a ferromagnet. The most plausible explanation for this effect is the appearance of pinning centers around 200 K, associated with a very early stage of the growth of the $\beta$ phase. With the aim of assessing unambiguously the appearance of phase coexistence and to give a quantitative description of its temperature evolution, we have investigated the magnetic behavior of MnAs epilayers grown over GaAs(100) and GaAs(111) by FMR experiments. This technique allows us to follow the changes of the ferromagnetic domain shapes with temperature that provide us necessary information to establish the temperature range of the phase coexistence.
%This technique is very sensitive and allows to measure very small changes in the magnetic behavior of a system, usually undetectable by magnetometric techniques. In particular, we were able to follow with accuracy the changes of the ferromagnetic domain shapes with temperature that provide us necessary information to establish the temperature range of the phase coexistence.

%\vspace{0.1in}
MnAs epilayers were grown by molecular beam epitaxy (MBE) as described elsewhere \cite{sterenHc}.
MnAs thin films of 66 nm thick were grown on (100) and (111) GaAs substrates, named M100-66 and M111-66
respectively. The ferromagnetic resonance measurements were performed at
$\nu$ $\sim$9.45 GHz (X-band).

%\vspace{0.1in}
\begin{figure}
\centering
\includegraphics[width=0.7\linewidth,angle=270]{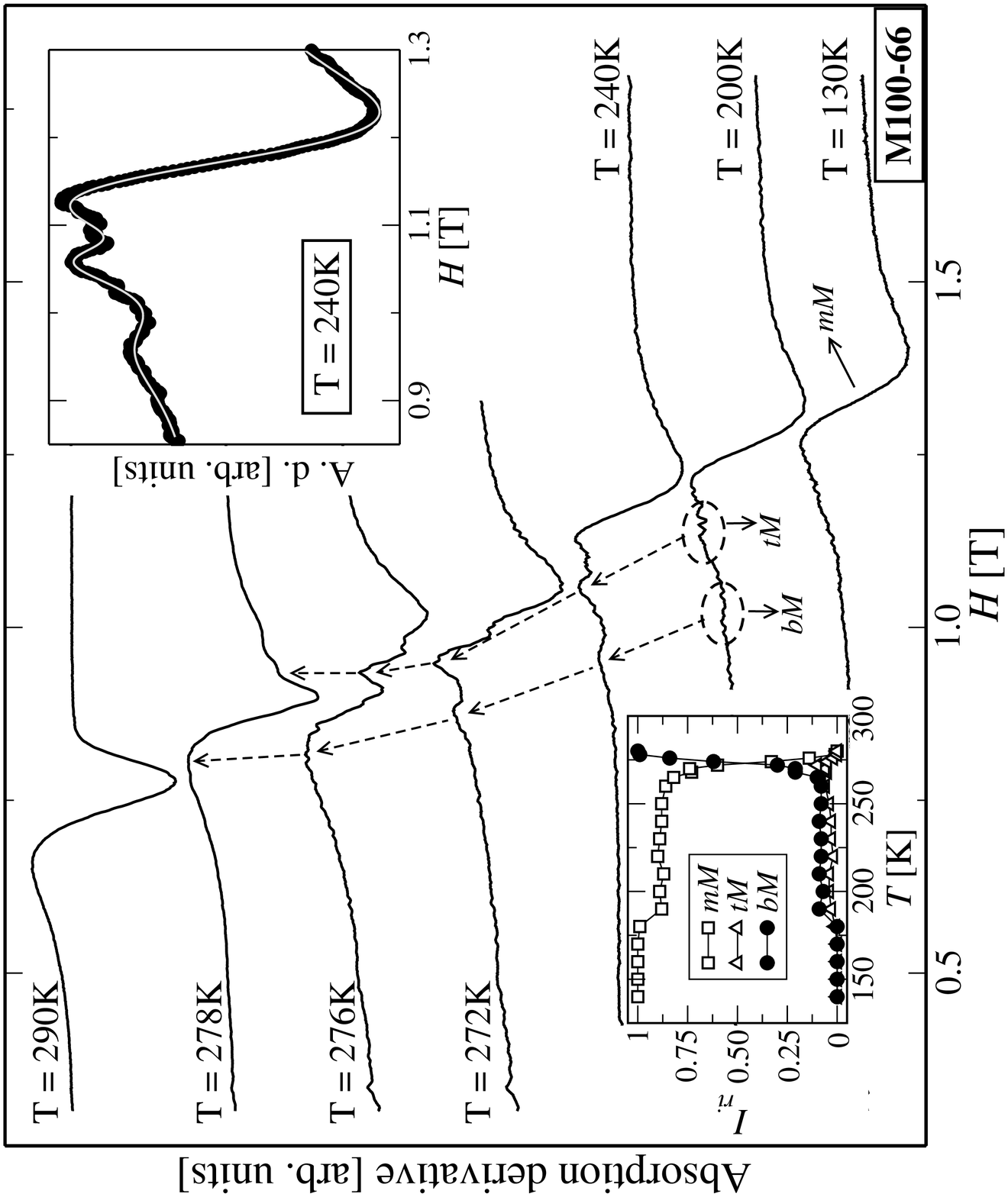}
\begin{minipage}{.23\textwidth}
\centering
\includegraphics[width=0.8\linewidth,angle=270]{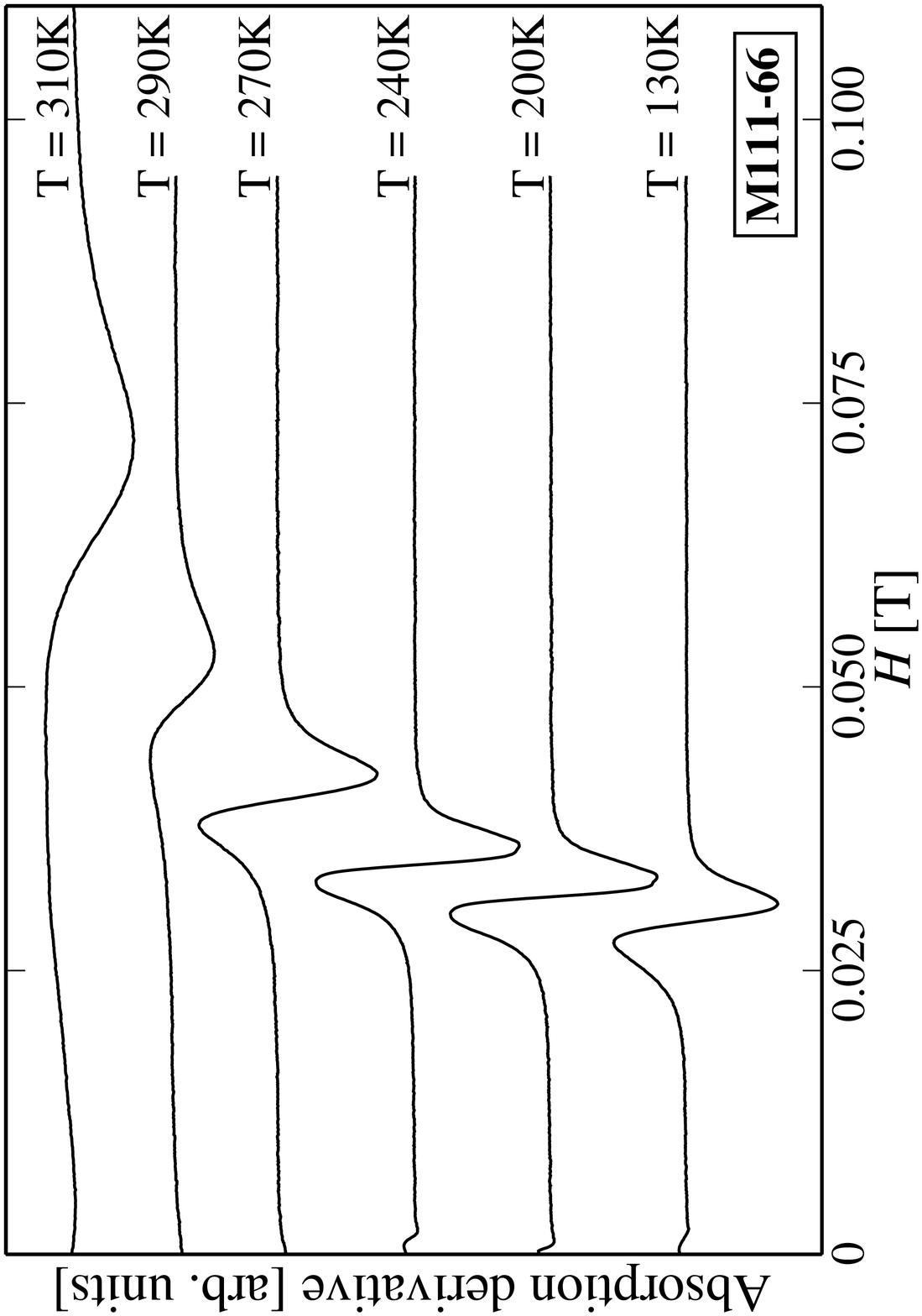}
\end{minipage}%
%\hspace{0.02\textwidth}%
\hfill
\begin{minipage}{.23\textwidth}
\centering
\includegraphics[width=0.775\linewidth]{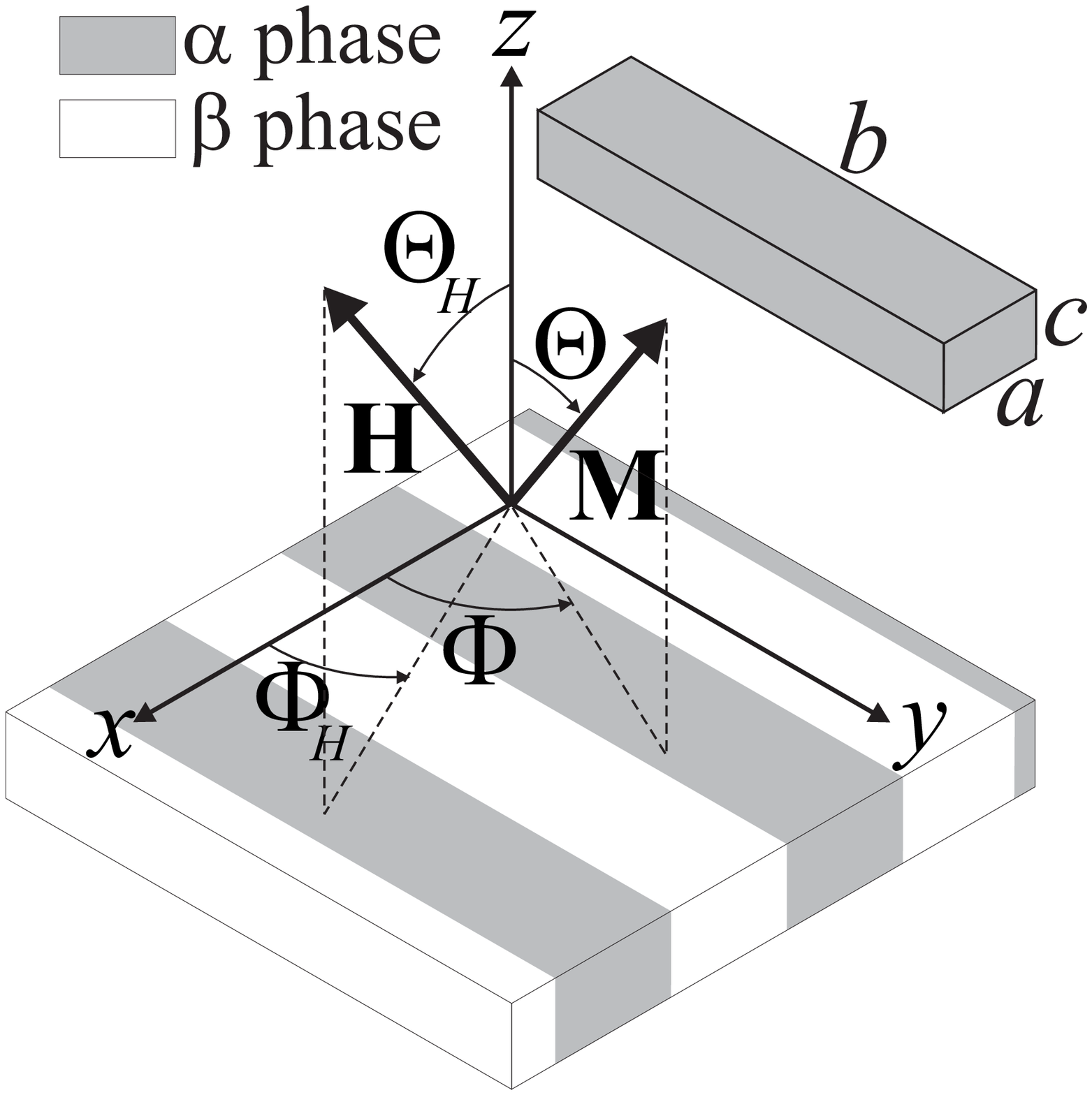}
\end{minipage}
\caption{\label{fig1} (a) FMR spectra in the out-of-plane geometry measured at different temperatures
for the M100-66 sample. The lower inset shows the intensity ratios of the
three modes, normalized to the total intensity of the
spectra. The upper inset shows the best fitted curve for the spectrum at T=240 K.
(b) FMR spectra in the in-plane geometry, measured at different temperatures
for the M111-66 sample. (c) Axis convention and scheme of the stripe-like domains in MnAs/GaAs(100) samples.}
\end{figure}
In Fig. \ref{fig1}(a), we show the FMR spectra corresponding to a
M100-66 film measured at different temperatures in the out-of-plane geometry, $\Theta_H=0$ [see Fig. \ref{fig1}(c)].
In the 130 K-272 K temperature range, we observe a main
resonance line, $mM$. At 180 K, the FMR spectra show two additional modes at
lower resonance fields, $bM$ and $tM$, respectively.
From 272 K to 278 K, the $mM$ and $tM$ modes become weaker and the intensity of the $bM$ mode
increases. The latter is the only one which is present at temperatures higher than 278 K.\\
Performing similar FMR experiments, A. Ney et al. \cite{ney} observed also two coexisting modes but only in a temperature range of 5 K (283K-288K) and they attribute this effect to the phase coexistance onset. However, our measurements show that the appearance of the additional modes occurs 100 K lower than the starting phase coexistance temperature.
%They probably did not observe the extra  modes at low temperature due to their small intensity and/or their large line-width. In other words, the appearance of the modes $tM$ and $bM$ indicates the onset of the $\beta$ phase at low temperature as we explained below.
%\begin{figure}
%\begin{center}
%\includegraphics[width=0.5\linewidth,angle=270]{MBE760_comp_lineas.eps}
%\end{center}
%\caption{\label{fig2}FMR spectra in the in-plane geometry, measured at different temperatures
%for the M111-66 sample.}
%\end{figure}\\
In order to quantify the fraction of the sample associated to each mode, we calculate the relative intensity of the resonance peaks. The modes intensity $I_i$ ($i$: $m$, $b$) is defined as the double integral of the FMR signal \cite{layadi} and is proportional to the magnetic moment of the sample fraction associated to each mode. We fit the FMR spectra with a sum of three Lorentzian curves. The relative intensity of each mode, $I_{ri}=I_i/I_t$ ($I_t$ stands for the total intensity), was adjusted in order to get the best fit of the whole spectra. In the inset of Fig. \ref{fig1}(a), we plot the temperature dependence of $I_{ri}$ for the three modes. Two remarkable features can be observed
from this figure: $i$) the intensity of the $tM$ and $bM$ modes
remain almost frozen from their onset and up to 272 K and $ii$) the temperature at which $I_{rb}$ increases abruptly agrees with the onset of the phase coexistence reported by the other authors \cite{ney,mattoso,vidal}.

In Fig. \ref{fig1}(b) we show the FMR spectra as a function of temperature for the M111-66 film in the in-plane geometry, $\Theta_H=\pi/2$. We observe a totally different behavior with respect to the other substrate orientation. All the FMR spectra present only one resonance mode in the whole temperature window of the study. We attribute this different behavior, with respect to that reported for M100-66 film, to a different topological arrangement of the $\alpha$ and $\beta$ domains during the phase coexistence. The absence of additional modes in the FMR spectra of M111-66 films is associated to an homogeneous distribution of the
$\alpha$ and the $\beta$ phases within the sample. In these conditions the FMR signal will not present changes compared to that of the pure $\alpha$ phase. These results agree with those obtained by other techniques \cite{takagaki,daweritzJVST}.

%\vspace{0.1in}
The existence of additional modes in the FMR spectra of the M100-66 films suggests that there are regions within the sample with different magnetic behavior than the usual $\alpha$ phase.
These regions can differ in their magnetization, anisotropy and/or
demagnetizing factor due to different domains shape. We discard changes in anisotropy during the phase coexistence based on Refs. \onlinecite{lindner,blois}.
In addition, we can neglect changes in magnetization, due to the fact that magneto-crystalline anisotropy is more sensitive to structural changes than magnetization; then we can expect that the magnetization of the ferromagnetic domains during the coexistence is similar to the bulk one as well.
Taking into account these aspects, we can assure that the shift
of the $tM$ and $bM$ modes compared to the $mM$ one is due to the existence of
zones with different demagnetizing factors. These results reveal that the phase coexistence starts self-ordered and that, the appearance of the $\beta$ phase drives the ferromagnetic domains to change their shapes in the new zones. We also observe that from 180 K to 278 K, the pure $\alpha$ phase regions and the zones where $\alpha$-$\beta$ phases coexist are present simultaneously in the sample.\\
The $\alpha$ and $\beta$ phase coexistence has been always modelled to be set in the whole sample from the onset temperature. This idea leads the authors of 
 Ref. \cite{kaganerPRL85} to propose as possible explanation of the phenomena that gliding misfit dislocations move along the MnAs/GaAs interface to rearrange the strains along sample. However, our results contradict partially this model. The coexistence of pure $\alpha$ phase and $\alpha$-$\beta$ phase regions below 278 K indicates that the strain spatial distribution is not homogeneous. Therefore, 
we suggest that the dislocations are not free to move along the interfaces due to the fact that their thermal energy is smaller than the pinning defects potentials at the intermediate temperature range.\\
Beyond 278 K, the fact that the peaks merge into a single mode indicates that above this temperature there is no more than one magnetic domains-type behavior in the samples, then the phase coexistence is extended along the whole sample.

%\vspace{0.1in}
In order to give a quantitative description of the magnetic behavior of the M100-66 sample, we model the different contributions to the free energy density, $F$, of our system. $F$ for the pure $\alpha$ phase region, associated to the $mM$ mode, is given by:
%\begin{figure}
%\centering
%\includegraphics[width=0.4\linewidth]{fmraxesplusstripesPRL.eps}
%\caption{\label{fig3}Axis convention and scheme of the stripe-like domains in MnAs/GaAs(100) samples.}
%\end{figure}
\begin{equation}
\label{eq1}
F_{mM} = - {\bf M}\cdot {\bf H}+\left( \frac{\mu_o}{2} \:N^{mM}_z \: M^2 + K_n \right) \cos^2 \Theta +K \sin^2 \Theta \: \sin^2 \Phi,
\end{equation}
where the first and second terms account for the Zeeman interaction and the demagnetizing contribution of a continuous thin film which lies in the  the $xy$ plane ($N^{mM}_z$=1) \cite{cullity}, respectively.
{\bf M} is the saturation magnetization and $K$ is the magneto-crystalline anisotropy for MnAs, which has a hard axis perpendicular to the hexagonal basal plane \cite{blois}. In our axis convention [Fig. \ref{fig1}(c)], the hard axis is parallel to the $y$ axis. $\Theta$ is the polar and $\Phi$ the azimuthal angle of the {\bf M}, respect to the [100] direction. $K_n$ is an uniaxial anisotropy,
that favors an in-plane magnetization and arises from the surface anisotropy and the intrinsic anisotropy caused by a slight distortion of the hexagonal base of the MnAs unit cell \cite{das,lindner}.\\
The energy density of the $\alpha$ phase in the coexistence zones, $F_{bM}$, takes into account the available
information about the magnetic domain shapes for T$>$273 K obtained from pictures of scanning tunneling and atomic force microscopies \cite{kastner,das}. Then, in order to model the self-ordered arrangement of the $\alpha$ and $\beta$ phases we will assume that the magnetic domains have a stripe-like shape for the whole temperature range. We consider that their shape is a rectangular prism and that side $b \gg a$,$c$ as shown in Fig. \ref{fig1}(c) \cite{cullity}. Then, $F_{bM}$ is expressed by \cite{lindner}, 
\begin{equation}
\begin{split}
\label{eq2}
F_{bM} =& - {\bf M}\cdot {\bf H}+\left( \frac{\mu_o}{2} \: N^{bM}_z \: M^2 + K_n \right) \cos^2 \Theta -\\
&\left( \frac{\mu_o}{2} \: N^{bM}_x \: M^2-K \right)\sin^2 \Theta \: \sin^2 \Phi,
\end{split}
\end{equation}
where $N^{bM}_x=N^{bM}_z-1$ \cite{cullity}. The free energy of the $tM$ mode is also described by an expression similar to the Eq. \eqref{eq2} but, with different demagnetizing parameters. 

%\vspace{0.1in}
\begin{figure}
\includegraphics[width=0.6\linewidth,angle=270]{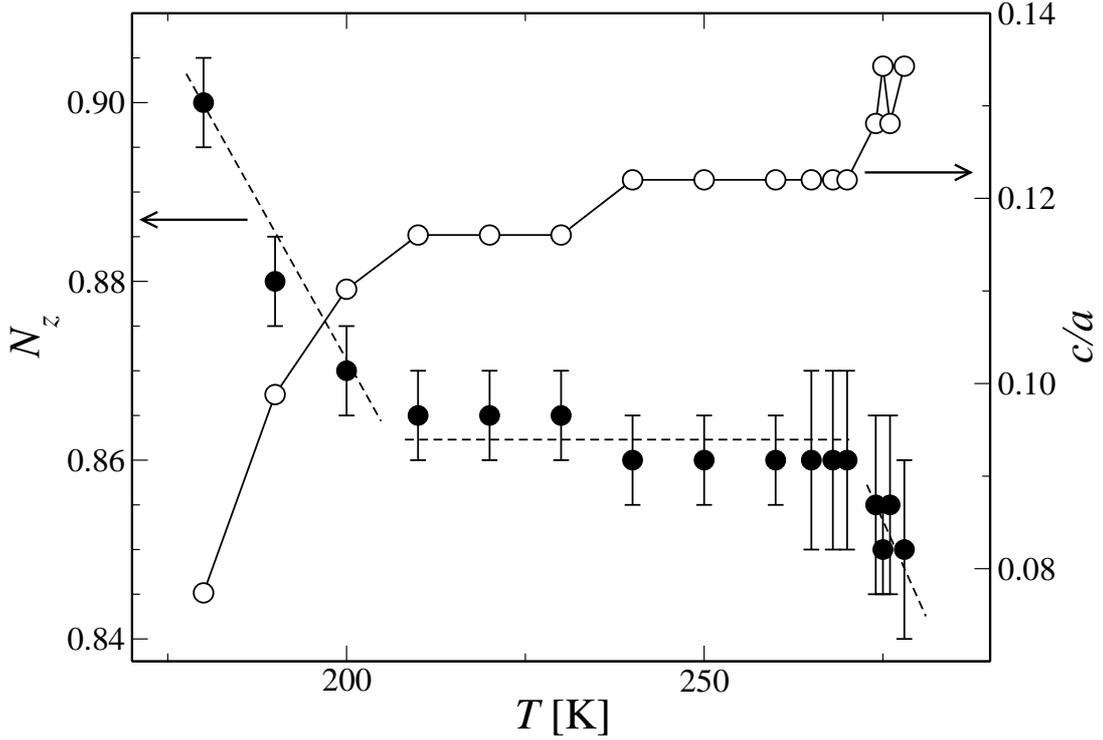}
\caption{\label{fig4}Demagnetization factor and $c$/$a$ ratio as a function of temperature.}
\end{figure}
To estimate the magnetic constants corresponding to our system, we perform a self-consistent calculation of the resonance fields using the equation derived by Ref. \onlinecite{smit}:
\begin{equation}
\label{eq3}
\omega^2=\frac{\gamma^2}{M^2\sin^2\Theta}\left[
\frac{\partial^2F}{\partial\Theta^2}\frac{\partial^2F}{\partial\Phi^2}-\left(
\frac{\partial^2F}{\partial\Theta\partial\Phi} \right)^2 \right],
\end{equation}
evaluated at the equilibrium angles $\Theta_{eq}$, $\Phi_{eq}$,
where $\gamma=g\mu_B/\hbar$ is the  gyromagnetic ratio. 
The equilibrium conditions for {\bf M}, $\Theta_{eq}$, $\Phi_{eq}$ are
obtained by minimizing the total free energy of the system ($F_{mM}+F_{bM}+F_{tM}$) for each orientation of the applied magnetic field. Taking into account only $F_{mM}$ we determine $K$ and $K_n$ by fitting the resonance field as a function of the polar angle, $\Theta_H$, and keeping $\Phi_H$ fixed (see Fig. \ref{fig1}(c) for each temperature; then,  we are able to determine the demagnetizing factor from $F_{bm}$ and $F_{tm}$. $M$ is obtained from magnetometric measurements.
The resonance field shift of the $bM$ mode in comparison with the $mM$ one is defined by the change of the demagnetizing factors; the thinner are the magnetic stripes, the larger is $N^{bM}_x$. In Fig \ref{fig4}, we show the temperature dependence of the calculated $N^{bM}_{z}$ value for the $bM$ mode. In this figure we can identify three different regimes: $i$) 180 K$<$$T$$<$200 K, $N^{bM}_{z}$ decreases from 0.90 to 0.865; $ii$) $N^{bM}_z$ remains almost constant up to 272 K and $iii$) for $T$$>$272 K, $N^{bM}_{z}$ decreases down to 0.85 at 278 K. We are unable to determine the $N^{bM}_z$ value beyond 278 K, because the resonance peaks merge into a single one. Lindner {\it et al.} \cite{lindner} deduced a demagnetizing factor of $N^{bM}_{z}$=0.84 for the stripes at room temperature.  
Our results indicate that for $T<$200 K and $T>$272 K the magnetic stripes become thinner [side $a$, as referred in Fig \ref{fig1}(c)]; and between 200 K and 272 K the stripes width remains almost constant.

Following Ref. \onlinecite{aharoni}, it is possible to determine the $c/a$ ratio from $N^{bM}_z$ for domains whose shapes are rectangular prisms. In Fig. \ref{fig4} the calculated $c/a$ ratio as a function of temperature is plotted. If we assume that the height (side $c$) of the stripes is the film thickness ($c$=66 nm) we find that the stripes width (side $a$) varies from $\sim$950 nm to $\sim$500 nm when we increase the temperature from 180 K to 278 K.
%We compare the calculated sizes with others obtained from STM images \cite{plake} in similar samples. These images were measured at room temperature and they are 150 nm and 440 nm width for 100 nm- and 200 nm-thick films, respectively. By doing a simple linear regression we estimate a value of 90 nm for the 66 nm-thick films. 
In the frame of this model, the higher resonance field of the $tM$ mode indicates that there are zones of ferromagnetic stripes of different sizes in the films, i.e. those associated to mode $tM$ are wider than those related to $bM$ mode. Due to the similarities of the ferromagnetic domains related to both, the $bM$ and $tM$ modes, we will only present the analysis of the data of the first mode. 

%\vspace{0.1in}
Summarizing, we investigated the coexistence of the $\alpha$ and $\beta$ phases in
MnAs thin films, grown over GaAs(100) and GaAs(111) substrates by performing
FMR experiments as a function of temperature and angles.
The spectra behavior
indicates that the phase coexistence is self-ordered for MnAs/GaAs(100) and it starts $\sim$100 K below the onset temperature detected by other experimental techniques. We show that the $\alpha + \beta$ regions are distributed along the samples in islands embedded in the $\alpha$ homogeneous phase for temperatures lower than 278 K. From 272 K, the volume of the $\alpha + \beta$ regions suffers an abrupt increase and, from 278 K, the coexistence is extended to the whole sample. The temperature which the strongest change is detected matches with the onset of phase coexistence reported in the literature. On the other hand, the FMR measurements in MnAs/GaAs(111) thin films agree with the fact that the $\beta$  phase inter-grows randomly within the $\alpha$ matrix.
%In addition to the better understanding of the magneto-crystalline transition in MnAs epilayers provided, this study opens a new stage where the determination of the arrangement of the $\alpha + \beta$ regions becomes important for a fully understanding of the role of the MnAs/GaAs interface in the magnetic properties of MnAs epilayers.

%\vspace{0.1in}
We acknowledge J. Pérez and R. Benávidez for technical assistance. The authors thank partial financial support from Fundaci\'on Antorchas, FONCyT  03-13297, Universidad Nacional de Cuyo, CONICET, ANR-MOMES and the ECOS-SUD. J.M and L.B.S are members of CONICET.

%\bibliography{refMnAs}
%\end{document}

\end{document}